\documentclass[pre,twocolumn,aps,superscriptaddress]{revtex4}
\usepackage[T1]{fontenc}
\usepackage{array}
\usepackage{booktabs}
\usepackage{mathrsfs}
\usepackage{mathbbol}
\usepackage{multirow}
\usepackage{amsmath}
\usepackage{amsthm}
\usepackage{amssymb}
\usepackage{stmaryrd}
\usepackage{graphicx}
\usepackage{latexsym}
\usepackage{textcomp}
\usepackage{mathtools}
\usepackage{xcolor}
\usepackage[utf8]{inputenc}
\usepackage[citecolor=magenta,colorlinks=true]{hyperref}
\usepackage{lineno}
\usepackage{stackrel}
\usepackage[toc,page,titletoc]{appendix}
\usepackage{bm}
\definecolor{mycolor1}{rgb}{0.1, 0.6, 0.6}

\begin{document}

\title{Influence of active breathing on rheology and jamming of amorphous solids: insights from microscopic and mesoscale analysis}

\author{Sayantan Ghosh\thanks{Sayantan Ghosh and Magali Le Goff contributed equally to this work.}}
\affiliation{The Institute of Mathematical Sciences, Taramani, Chennai 600113, India}

\author{Magali Le Goff\footnotemark[1]}
\affiliation{Université Grenoble Alpes, CNRS, LIPhy, 38000 Grenoble, France}

\author{Pinaki Chaudhuri}
\affiliation{The Institute of Mathematical Sciences, Taramani, Chennai 600113, India}

\author{Kirsten Martens}
\email{kirsten.martens@univ-grenoble-alpes.fr}
\affiliation{Université Grenoble Alpes, CNRS, LIPhy, 38000 Grenoble, France}

\date{~\today}

\begin{abstract}
We study the flow behavior and unjamming transition in dense assemblies of actively deforming particles that periodically change size, a process that we refer to as breathing. Using large-scale molecular dynamics simulations and a complementary mesoscale elasto-plastic model, we explore how this internal activity influences plasticity and rheology. At low amplitudes of breathing, the system remains jammed and displays localized, reversible rearrangements. As the amplitude of the breathing increases beyond a critical threshold, the system undergoes an activity-induced fluidization marked by a surge in plastic events and a drop in yield stress. The flow curve analysis reveals a transition from yield-stress behavior to Newtonian flow at high activity. The mesoscale model captures these trends and provides insight into the role of stress redistribution due to local active deformations. Our findings highlight the potential of internal active driving to tune the mechanical state of amorphous materials without external forcing.
\end{abstract}

\maketitle

\section{Introduction}

Active matter systems consist of units that continuously consume energy to generate motion or internal stresses, thereby driving the system far from thermodynamic equilibrium \cite{te2025metareview}. In contrast to externally driven materials, such as sheared suspensions, active systems are internally powered at the microscopic scale, giving rise to rich and often unexpected collective behaviors \cite{marchetti2013hydrodynamics,vicsek2012collective}. Examples span a broad range of biological and synthetic systems, including bacterial suspensions \cite{copeland2008bacterial}, self-propelled colloids \cite{theurkauff2012dynamic,buttinoni2013dynamical}, and confluent cell monolayers \cite{poujade2007collective,mehes2014collective,angelini2011glass}.

A striking case is found in dense epithelial tissues, where cells display fluid-like collective dynamics despite strong crowding.\cite{angelini2011glass}. These collective dynamics are driven by multiple sources of cellular activity, including: (i) directed motility such as crawling \cite{rafelski2004crawling, sugino2024non}; (ii) active mechanical deformations through protrusion, contraction, and cell shape control \cite{zehnder2015cell,taloni2015volume,bi2015density,PhysRevLett.130.058202,lawson2021jamming}; and (iii) cell turnover via division and apoptosis \cite{ranft2010fluidization,matoz2017nonlinear,matoz2017cell}. While the role of motility has been extensively modeled using self-propelled particle approaches \cite{vicsek1995novel,henkes2011active,wiese2023fluid}, less attention has been devoted to understanding the mechanical consequences of cell size fluctuations \cite{tjhung2017discontinuous,li2025fluidization}, which are prominently observed in experimental studies of cell assemblies. Epithelial cells for example (e.g., MDCK cells) exhibit spontaneous, ATP-dependent oscillations in their volume, even in confluent tissues.\cite{zehnder2015cell}.

To explore this mechanism in isolation, we model a system of soft particles that undergo periodic size changes, a process we term active breathing. Previous studies have shown that such internal deformations, even in the absence of self-propulsion, can fluidize an otherwise jammed system by triggering localized, irreversible rearrangements \cite{tjhung2017discontinuous}. The resulting transition from an arrested to a flowing state bears resemblance to the yielding behavior observed in amorphous solids, where flow emerges once a critical stress threshold is exceeded \cite{bonn2015yield,knowlton2014microscopic,jaiswal2016mechanical}. However, in the case of active breathing, the transition is driven purely by internal dynamics.

In this work, we combine large-scale particle-based simulations with a mesoscale tensorial elasto-plastic model to investigate how active breathing modulates the mechanical response. We first characterize the spontaneous unjamming that occurs above a critical activity threshold, then examine the rheological behavior of the system under imposed shear. The observed activity-dependent reduction in yield stress and the crossover from shear-thinning to Newtonian flow are qualitatively reproduced by the mesoscale model, which also captures the spatial structure of stress propagation and localization around active sites.

These results demonstrate that active size fluctuations can serve as a powerful internal control parameter for tuning flow behavior in jammed materials. They offer new insights into how mechanical regulation can emerge intrinsically in both biological and synthetic active matter systems.

\section{Model and Methods}

\subsection{Microscopic model}

For our study, we consider a two-dimensional 60:40 binary mixture with $N=16000$ particles, with initial diameters $0.71d_{0}$ and $d_{0}$ respectively. We work at a packing fraction of 0.94, i.e. the system is jammed in the passive state~\cite{heussinger2009jamming}. Similar to the previous study \cite{tjhung2017discontinuous}, the particles interact via a harmonic potential:
\begin{equation}
 U(r_{ij})=\frac{\epsilon}{2}(1-\frac{r_{ij}}{\sigma_{ij}})^{2}H(d_{ij}-r_{ij})
\end{equation}
where $H(x\geq0)=1$, $r_{ij}=\lvert\vec{r_{i}}-\vec{r_{j}}\rvert$, $d_{ij}=(d_{i}+d_{j})/2$, with $d_{i}$ and $\vec{r_{i}}$ being the diameter and position of the particle $i$ respectively.

Active breathing occurs via the oscillation of the diameter of the particle, as given by:
\begin{equation}
 d_{i}(t)=d_{i0}(1+a\cos(\omega t+\psi_{i}))
\end{equation}
where $T=2\pi/\omega$ is the period of oscillation. We have chosen $T=820\tau_{0}$~\cite{tjhung2017discontinuous}, implying that the oscillations are slow. $\psi_{i}$ is a phase factor chosen in a way that it keeps the packing fraction constant, viz. $\psi_{i}=2\pi i/N$ where $N$ is the number of particles with the same initial diameter. 

We investigate the dynamical behaviour of the breathing particles, under athermal conditions, using Langevin dissipative dynamics:
\begin{equation}
 m\frac{d^{2}\vec{r_{i}}}{dt^{2}}+\xi\frac{d\vec{r_{i}}}{dt}=-\sum_{j\neq i}^{} \frac{\partial V(r_{ij})}{\partial\vec{r_{j}}}
 \label{eom}
\end{equation}
where $\xi$ is a friction coefficient. The dissipation timescale is given by $\tau_{0}=\xi\sigma_{0}^{2}/\epsilon$, and we choose $\tau_{0}=1$.
The equations of motions are numerically integrated using a suitably modified version of LAMMPS \cite{thompson2022lammps}.

For studying the response to applied shear, we consider two different protocols. At first, we study the response to imposed shear-rate $\dot{\gamma}$, whereby the simulation box is deformed at some rate defined by the imposed $\dot{\gamma}$ and the particles react to such deformation via Eqn.\ref{eom}; in this study, we gather the steady-state rheological information arising out of the shear response. Subsequently, we study the transient response to applied shear-stress ($\sigma_0$), using a recently devised feedback method wherein we have an additional equation of motion for box deformation \cite{cabriolu2019precursors}. Both these protocols are implemented using LAMMPS.

\subsection{Mesoscale model}

To mimic the dynamics observed in our microscopic simulations, we develop a two-dimensional tensorial elasto-plastic model (EPM) that incorporates local active driving \cite{legoff:tel-03426944}. This approach builds on the framework of elastoplastic models for amorphous solids \cite{review}, and has recently also been adapted to describe active materials such as active glassy systems under internal forcing in form of self-propulsion \cite{ghosh2025elastoplastic}. To simulate the dynamics of a dense system of particles deforming in the form of oscillations of their radius, we implement in our active elasto-plastic model (AEPM) the local driving under the form of the elastic response to point-like isotropically deforming particles as predicted by linear elasticity, detailed in appendix section \ref{app:active_response}.

\begin{figure}
    \centering
    \includegraphics[width = 0.9\columnwidth, clip]{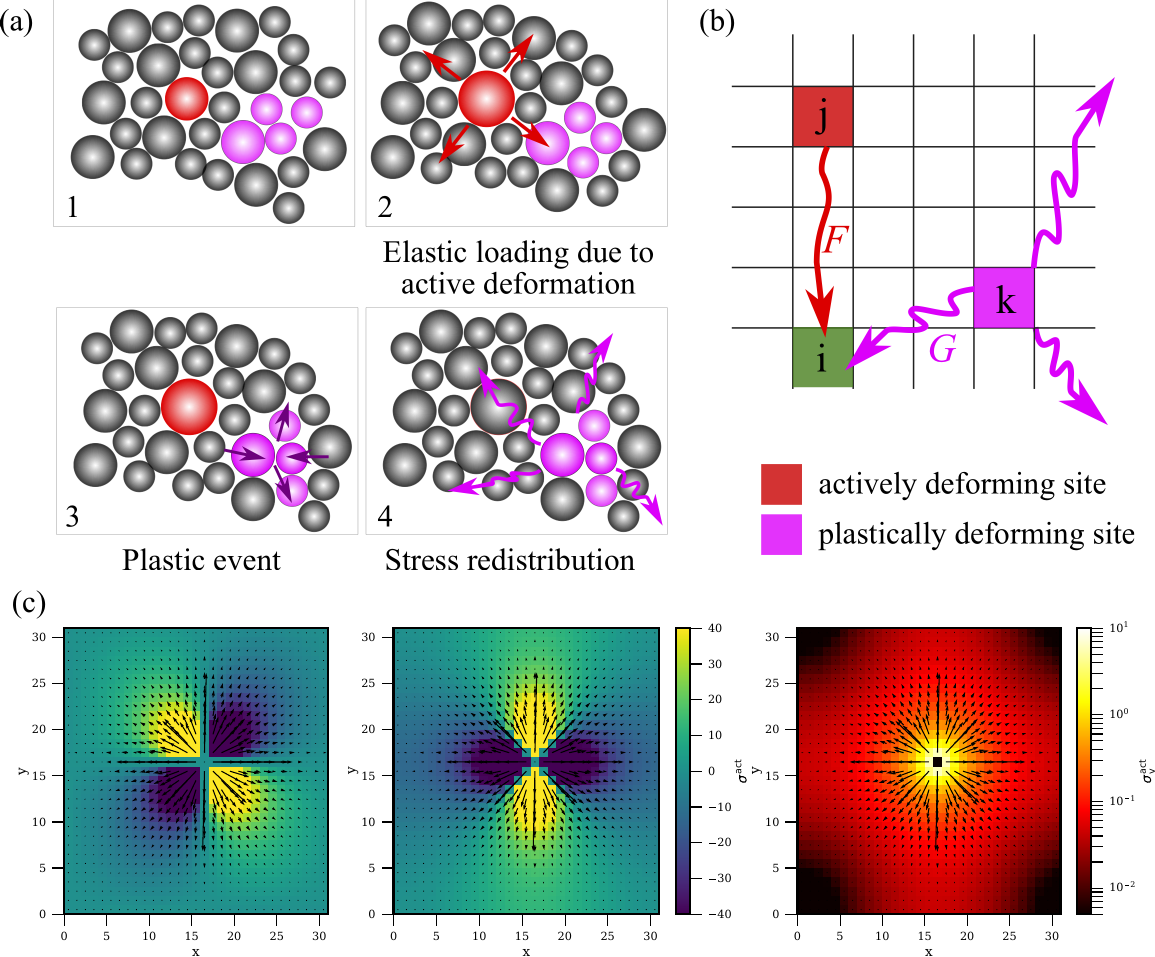}
    \caption[Principle of the mesoscale active elasto-plastic model]{\textbf{Principle of the mesoscale active elasto-plastic model}. (a) Sketch of plastic deformation induced by an internal activity. 1 and 2: a particle (in red) undergoing an active deformation induces a displacement of the surrounding particles and elastic stress in the surrounding material. 3: This increase in stress induces a plastic rearrangement of particles (magenta particles). 4: This plastic rearrangement leads to a stress redistribution in the system. (b) \textbf{Lattice model describing the active elasto-plastic scenario}: the stress state at site $i$ (green) results from the contribution of the actively deforming sites (red), described with an elastic propagator $F$ and from the plastically deforming sites (magenta), described with an Eshelby-like elastic propagator $G$.
    (c) \textbf{Snapshots of stress and displacement fields in the AEPM} for a single actively deforming site in the center of the system without enabling for relaxation by plastic events (the stress scale and the displacement vectors scale are chosen for visualization purpose). Left: $\sigma_{xy}^\mathrm{act}$ component. Middle: $\sigma_{xx}^\mathrm{act}$ component. Right: von Mises stress $\sigma_\mathrm{v}^\mathrm{act}$}. 
    \label{fig:AEPM_principe}
\end{figure}

We consider a two-dimensional elasto-plastic tensorial model with a local driving.
The driving is implemented under the form of an imposed local stress that depends upon the distribution of actively deforming sites (red sites in Fig.~\ref{fig:AEPM_principe}(b)).

The spirit of this model is the same as the standard elasto-plastic models developed for simple shear \cite{review}, except that instead of considering a homogeneous drive and applying the same value $\sigma^{\mathrm{ext}}$ to all sites, we define a new stress field, $\sigma^{\mathrm{act}}$, that results from the local contraction or dilation of particles.
This stress field $\sigma^{\mathrm{act}}$ is computed as the convolution of an active deformation field $\gamma^{\mathrm{act}}$ with an elastic propagator $F$, as depicted in Fig.~\ref{fig:AEPM_principe}(b).
We consider here only the case of purely dilating or contracting inclusions.

%%%%%%%%%%%%%%%%%%%%%%%%%%%%%%%%%%%%%%%%%%%%%%
%%%%%%%%%%%%%%%%%%%%%%%%%%%%%%%%%%%%%%%%%%%%%%%

In our mesoscale model, we further assume that the medium is incompressible, and the stress response reads, in Fourier space:

\begin{equation}
    \Tilde{\sigma}^{\mathrm{act}}_{xy}(q_x,q_y) =  \mu \frac{2q_x q_y}{q^2} \Tilde{\gamma}^\mathrm{act}
\end{equation}
\begin{equation}
    \Tilde{\sigma}^{\mathrm{act}}_{xx}(q_x,q_y) = - \Tilde{\sigma}^{\mathrm{act}}_{yy} = \mu \frac{q_x^2-q_y^2}{q^2} \Tilde{\gamma}^\mathrm{act}
\end{equation}
for $q_x \neq 0$, $q_y \neq 0$ and- $\Tilde{\sigma}^{\mathrm{act}}_{xy}(0,0) = \Tilde{\sigma}^{\mathrm{act}}_{xy}(0,0) = 0$, with $\mu$ the shear modulus and $\Tilde{\gamma}^\mathrm{act}$ the Fourier transform of the active deformation field.
We show the stress components $\Tilde{\sigma}^{\mathrm{act}}_{xy}$ and $\Tilde{\sigma}^{\mathrm{act}}_{xx}$ in response to a single actively deforming (dilating) site located in the center of the system in Fig.~\ref{fig:AEPM_principe}(c, left and middle panels).
For simplicity we set the value at the origin to 0. This assumption is not very realistic due to the fact that a mesoscale active block includes in reality several neighboring particles in addition to the active particle and is thus subjected to stress redistribution. 
Testing more realistic rules for the local stress response should be addressed in future work.
For simplicity we will focus here on this first version of the AEPM and test whether it gives the same qualitative behavior as the particle-based simulations.

The local stress results from three contributions:
\begin{equation}
    \sigma_{\alpha\beta}(\vec{x},t) =  \sigma_{\alpha\beta}^{\mathrm{ext}}(t) + \sigma_{\alpha\beta}^{\mathrm{act}}(\vec{x},t) + \sigma_{\alpha\beta}^{\mathrm{int}}(\vec{x},t)
\end{equation}
where $\sigma_{\alpha\beta}^{\mathrm{ext}}$ is an externally imposed stress (e.g., applied shear), $\sigma_{\alpha\beta}^{\mathrm{act}}$ results from the internal active driving and $\sigma_{\alpha\beta}^{\mathrm{int}}$ describes the stress redistribution due to localized plastic events.
Using a shear-rate-controlled protocol, the dynamics of the stress reads:
\begin{eqnarray}
    \frac{\partial \sigma_{\alpha\beta} (\vec{x},t)}{\partial t} &=& \mu \dot{\gamma}(t) + \mu \int d^d\vec{x'}  G_{\alpha\beta,\gamma\delta}(\vec{x}-\vec{x'}) \frac{\partial \gamma^{\text{pl}}_{\gamma\delta}(\vec{x'},t)}{\partial t}  \nonumber \\
    &&+ \mu \int d^d\vec{x'}  F_{\alpha\beta}(\vec{x}-\vec{x'}) \frac{\partial \gamma^{\text{act}}(\vec{x'},t)}{\partial t}
\end{eqnarray}
with $G$ the usual two-dimensional Eshelby kernel \cite{review},
$F$ an elastic kernel describing the response to a dilating or contracting inclusion,
$\dot{\gamma}(t)$ an externally imposed shear rate, $\gamma^{\text{pl}}$ the plastic deformation field and $\gamma^{\text{act}}$ the active deformation field.
The dynamics of the plastic deformation $\gamma^{\text{pl}}_{\alpha\beta}$ remains the same as in the standard elasto-plastic models \cite{review}
\begin{equation}
 \frac{\partial\gamma^{\text{pl}}_{\alpha\beta}(\vec{x},t)}{\partial t} = \frac{n(\vec{x},t)\sigma_{\alpha\beta}(\vec{x},t)}{\mu\tau}
\end{equation}
with $n(\vec{x},t)$ the local plastic state which has its own dynamics, determined by a yielding rule based on a yield stress distribution and a recovery rule with a fixed rate.

We rewrite the above expression with discretized spatial coordinates ($i,j$). 

\begin{eqnarray}
\frac{d}{dt}\sigma_{\alpha\beta}(i,j)&=&\mu\dot{\gamma}+\mu\underset{i'j'}{\sum}G_{\alpha\beta,\gamma\delta}\frac{d}{dt}\gamma^{\mathrm{pl}}_{\gamma\delta}(i',j') \nonumber \\
&& +\mu\underset{i'j'}{\sum}F_{\alpha\beta}\frac{d}{dt}\gamma^\mathrm{act}(i',j')
\label{eq_active}
\end{eqnarray}
with $\gamma^\mathrm{act}(i',j')$ the active deformation undergone by the site $(i',j')$, related to radius change of an active particle:
$\gamma^\mathrm{act} \sim \Delta r/r_0$.
The plastic deformation $\gamma^\mathrm{pl}_{\alpha\beta}$ reads
\begin{equation}
    \frac{d \gamma_{\alpha\beta}^{\mathrm{pl}}(i,j)}{d t} = 
    n(i,j) \frac{\sigma_{\alpha\beta}(i,j)}{\mu \tau}
\end{equation}
\noindent with $\tau$ a relaxation timescale and $n(i,j)$ the plastic activity at site $(i,j)$.
A site becomes plastic when the local yielding criterion is met, using a von Mises criterion $\sigma_\mathrm{v}(i,j) > \sigma_\mathrm{y}$, with the local threshold $\sigma_\mathrm{y}$ drawn from a distribution and renewed after each yielding event.
When a site has yielded, it becomes elastic again after a typical time $\tau_\mathrm{el}$.

We use an incompressibility assumption, considering two components of the stress tensor: $\sigma_{xx}=-\sigma_{yy}=(\sigma_{xx}-\sigma_{yy})/2$ and $\sigma_{xy}$.
In this case, the local yielding criterion is $\sigma_\mathrm{v}(i,j)=\sqrt{\sigma_{xx}^2(i,j)+\sigma_{xy}^2(i,j)} > \sigma_\mathrm{y}$.
The propagators $G$ and $F$ thus read, in Fourier space:
\begin{equation}
   \Tilde{G} = \frac{1}{q^4} \begin{bmatrix}
-(q_x^2-q_y^2)^2 & -2q_x q_y(q_x^2-q_y^2) \\
-2q_x q_y(q_x^2-q_y^2) & -4q_x^2q_y^2
\end{bmatrix}
\label{eq:propagator_tensorial_eshelby}
\end{equation}
\begin{equation}
   \Tilde{F} = \frac{1}{q^2} \begin{bmatrix}
2 q_x q_y  \\
q_x^2 - q_y^2
\end{bmatrix}
\label{eq:propagator_tensorial_active}
\end{equation}

We use a mesh refinement \cite{nicolas2013mesoscopic} to compute the convolution in Fourier space both for the Eshelby contribution and for the active contribution to the stress.

\begin{figure*}
  \includegraphics[scale=0.88]{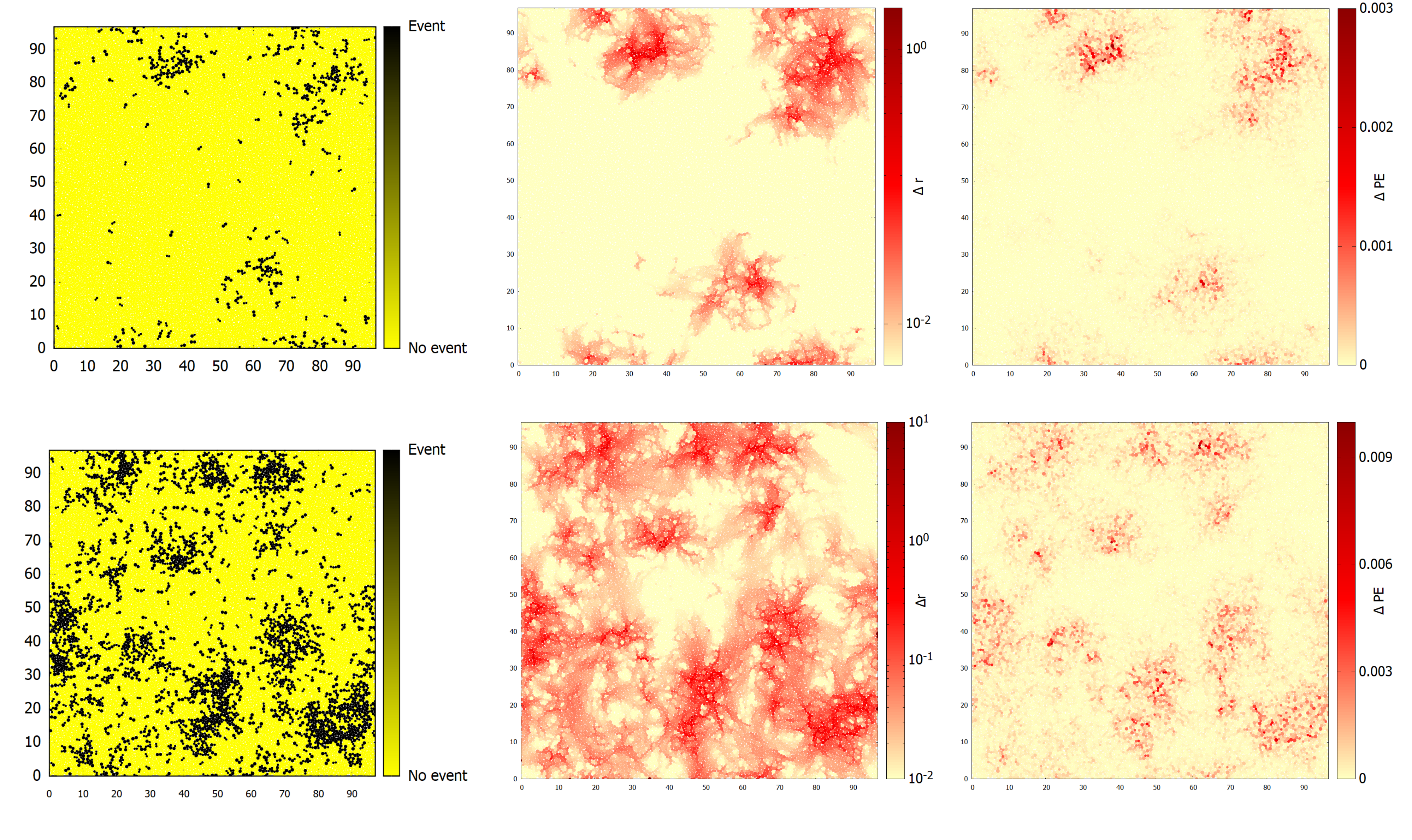}
\caption{{\bf Microscopic model.} For two consecutive stroboscopic frames at zero strain,  spatial maps of (left column) change in local contacts. (middle column) single particle root mean squared displacements, (right column) difference in local potential energy. Top and bottom panel correspond to $a=0.049, 0.051$ respectively.}
\label{f0}
\end{figure*}

\begin{figure*}
   \centering
  \includegraphics[scale=0.36]{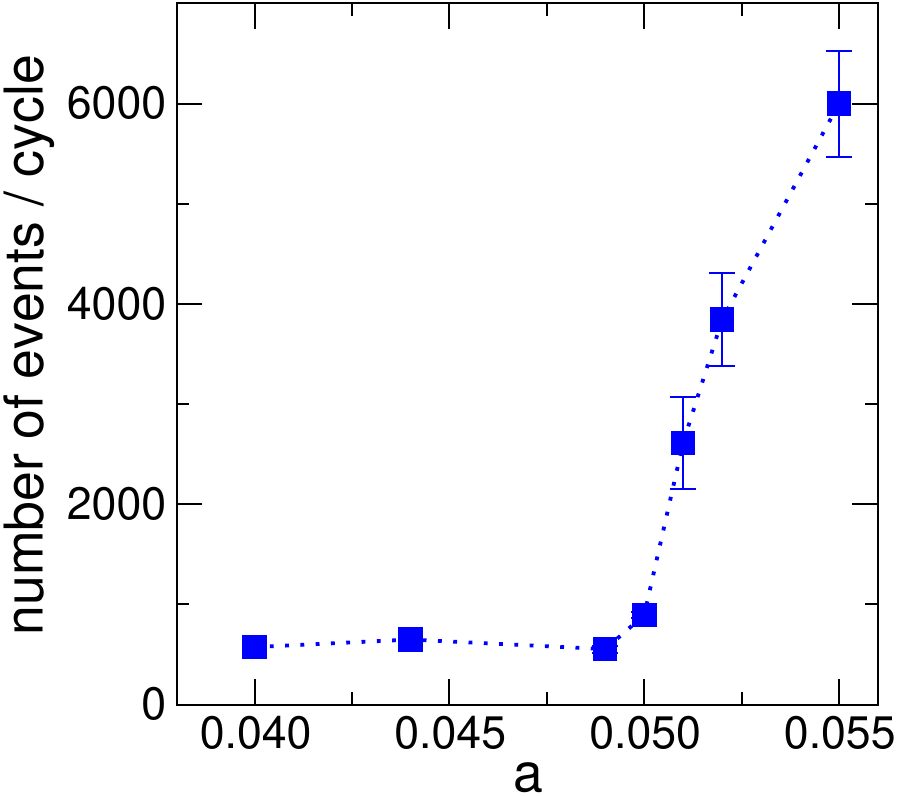}
    \includegraphics[scale=0.4]{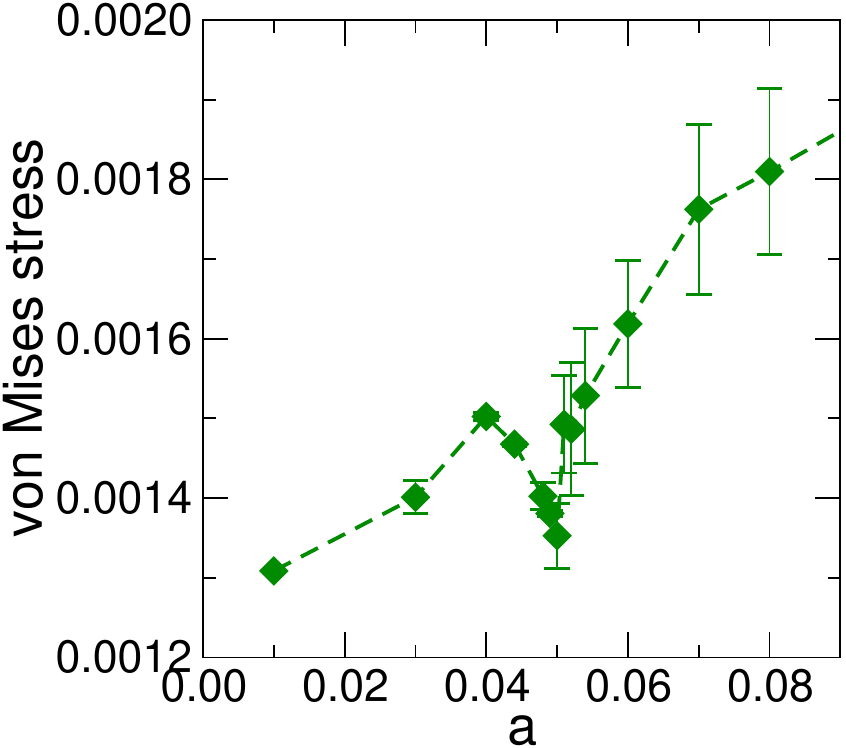}\\
    \includegraphics[scale=0.4]{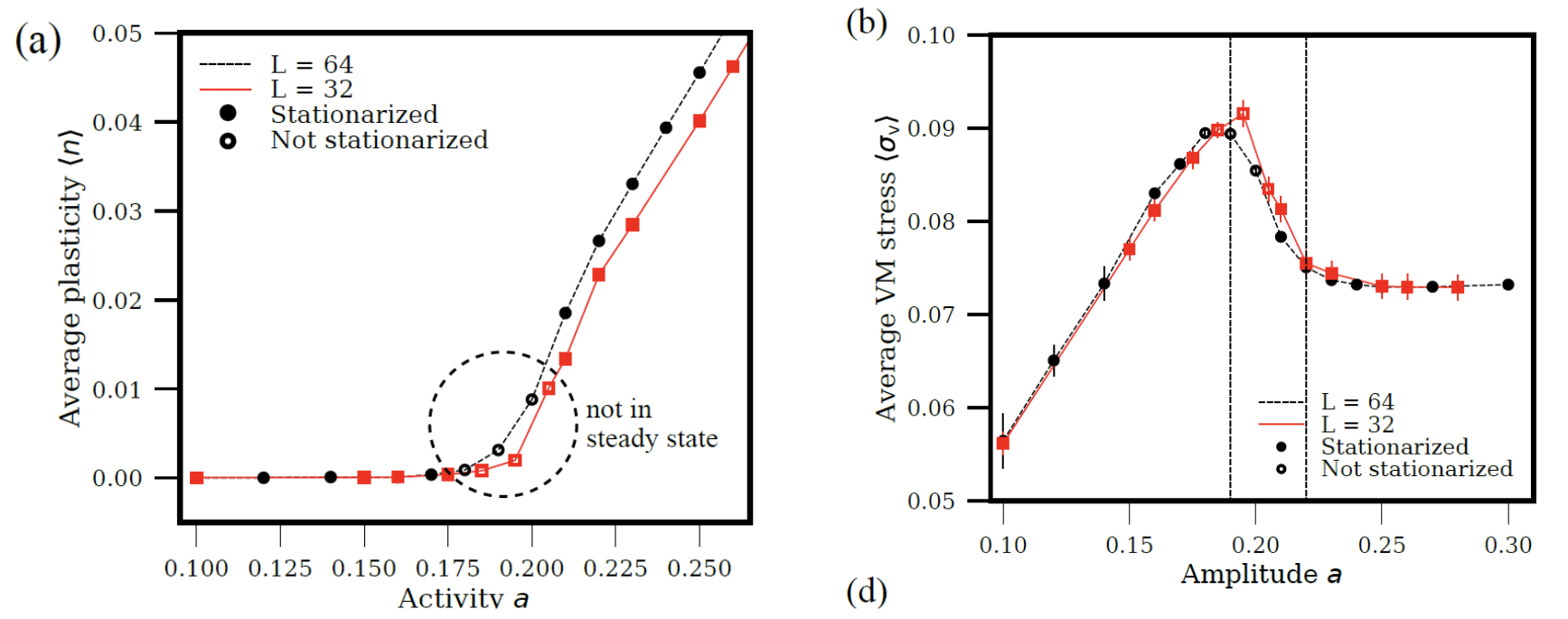}
    \caption{(Top). {\bf Microscopic model.} (left) Variation of number of events per cycle with activity. The error bars are shown. (right) von Mises stress plotted against activity. The error bars are shown.  (Bottom). {\bf Mesoscale model.} (left) Variation of number of events per cycle with activity. The error bars are shown. (right) von Mises stress plotted against activity. The error bars are shown. }
    \label{fig3}
\end{figure*}

\section{Results}

\subsection{In the absence of external drive}

%\subsubsection{Microscopic Model}

\subsubsection{Defining a plastic event}

Within the microscopic simulations, to develop the phenomenology similar to the elastoplastic model where local active sites are intrinsic descriptors, we need a strategy to define a plastic event at the particulate scale. A possible way to define such 
an event is by identifying whether any of the 
constituent particles changes neighbours in between two frames -- since the inter-particle interaction become finite only on contact, any contact breaking in this dense limit can be deemed as a change of neighbours. Alternative definitions correspond to events being locations of large mean squared displacement experienced by a single particle between the two
frames, or large changes in potential energy between the two configurations of interest. We show in Fig.\ref{f0} that all these definitions give similar information, viz. the location of a contact break is also the location of large mean squared displacement as well as large change in local potential energy. 
For ease of calculation, we use the change of contact for defining the occurrence of an event.

In the top and bottom panels of Fig.\ref{f0}, we show data for values of $a$, which are slightly larger or smaller than the threshold reported in Ref.~\cite{tjhung2017discontinuous}. We observe that the number of events between two stroboscopic frames distinctly increases with increase in breathing amplitude. We systematically calculate the steady-state average of the number of such events occurring between the beginning and the end of the cycle; the data is shown in the left-top panel of Fig.~\ref{fig3}. We observe that the number of such events remains constant until a certain value of $a$ and then suddenly starts growing. The approximate threshold for such increased proliferation of events per cycle is around $a_c$ reported by Ref.~\cite{tjhung2017discontinuous}. Within the scope of the mesoscale model, a similar increase in plasticity, quantified by the number of active sites per cycle, is also observed; see left-bottom panel of Fig.~\ref{fig3}.

\subsubsection{Measuring stresses}

Since the loss of rigidity of the material at some threshold breathing amplitude is akin to yielding, we would like to track the evolution
of the stress that develops in the system under steady state conditions, for increasing $a$. Since a breathing event involves the development of different stress components as discussed above,
we therefore consider the von Mises stress to check the mechanical state of the system. The von Mises stress for the two-dimensional system is defined as:
\begin{equation}
    \sigma_{v}=\sqrt{(\frac{\sigma_{xx}-\sigma_{yy}}{2})^{2}+\sigma_{xy}^{2}}
\end{equation}
The data from the microscale simulations are shown in the top-right panel of Fig.\ref{fig3}. We observe that around the $a$ value where the number
of events per cycle starts increasing, the von Mises stress goes through a cusp. A similar non-monotonicity, via an overshoot, is observed in the data for the mesoscale model; see bottom-right panel of Fig.\ref{fig3}.

%\begin{figure*}[h]
%    \centering
%    \includegraphics[scale=0.9]{Picture3.png}
%    \caption{{\bf Microscopic model.} }
%    \label{fig:3}
%\end{figure*}

%\begin{figure*}[h]
%    \centering
%    \includegraphics[scale=0.9]{Picture4.png}
%    \caption{{\bf Microscopic model.} }
%    \label{fig:4}
%\end{figure*}

\begin{figure*}
%    \centering
    \includegraphics[scale=0.4]{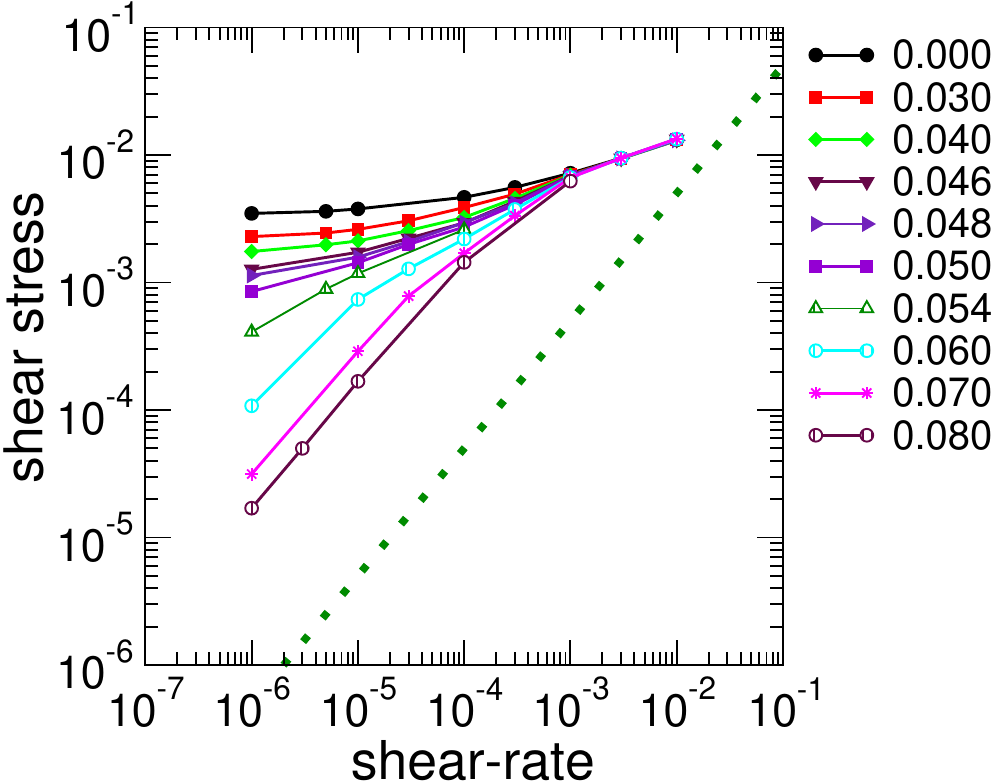}
   \includegraphics[scale=0.4]{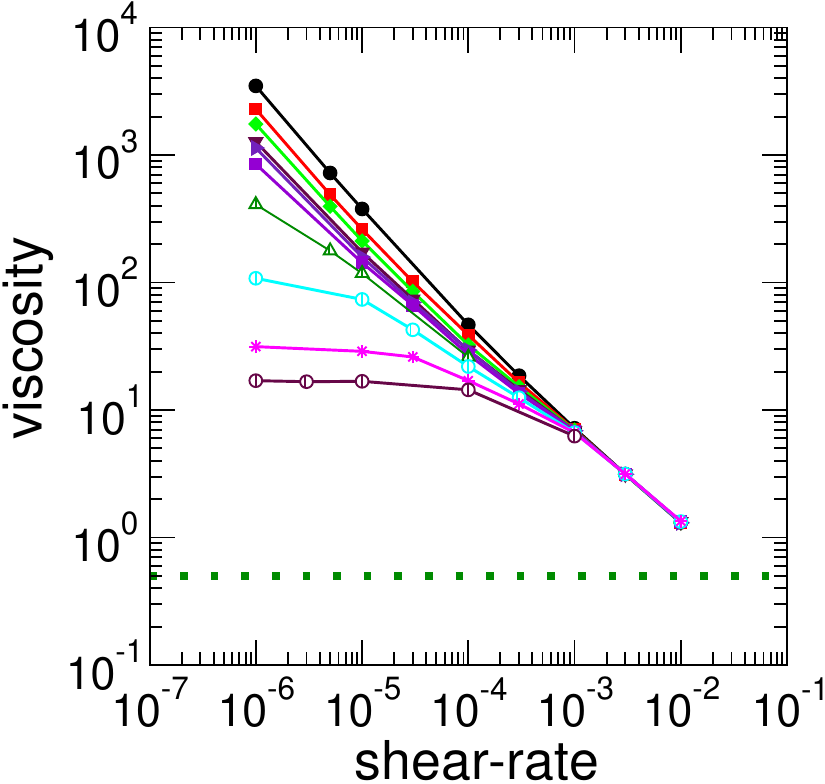}\\
\includegraphics[scale=0.47]{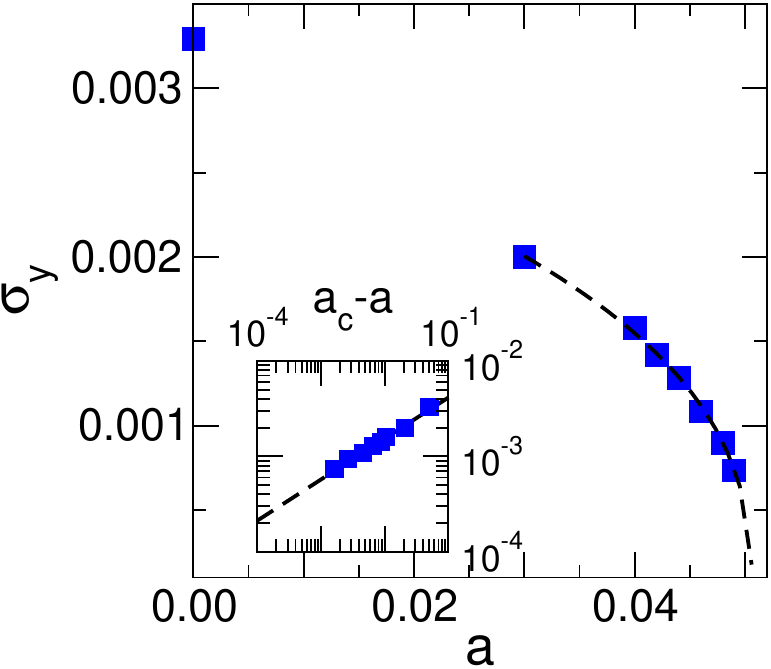}
\includegraphics[scale=0.5]{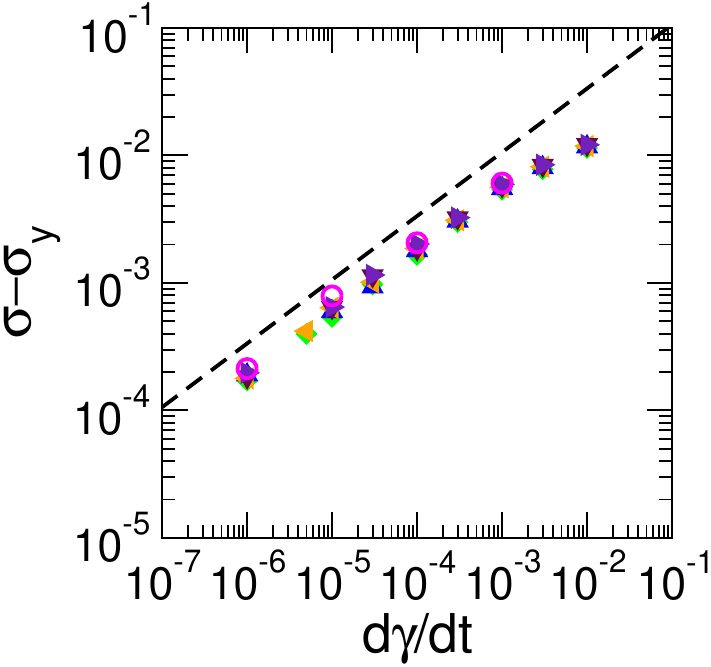}
\caption{{\bf Microscopic model.} Rheological data: [Top panel.] (Left)  $\sigma$ vs $\dot{\gamma}$ for different amplitudes of breathing, as marked. (Right) Corresponding shear-rate dependent viscosity $\eta=\sigma/\dot{\gamma}$. [Bottom Panel.]  (Left) Variation of the yield stress, $\sigma_y$, with breathing amplitude $a$ as obtained from Herschel-Bulkley fits to $\sigma$ vs $\dot{\gamma}$ data. Dashed line shows a fit with a power-law $(a_c-a)^\zeta$ which estimates a rigidity threshold of $a_c=0.0506$ with $\zeta \approx 0.39$. The inset exemplifies the power-law fit to the data, for $\sigma_y$ vs $(a_c-a)$. (Right) $\sigma - \sigma_y$ vs $\dot{\gamma}$ for flow curves for amplitudes $a < a_c$, showing reasonable collapse indicating similar shapes;
dashed line corresponds to a power-law with exponent 1/2).}
    \label{fig:6}
\end{figure*}

\subsection{Response to applied shear}

\subsubsection{Microscopic model}

Next we probe the response of the athermal assembly of breathing particles to macroscopic deformation via an imposed shear-rate $\dot{\gamma}$. Initially we prepare the steady state configurations corresponding to a breathing amplitude $a$ and subsequently impose the external shear. For every such $\{a,\dot{\gamma}\}$, we obtain the steadily flowing state and record the average shear-stress ($\sigma_{xy}$) that has developed. The flow curve (variation of $\sigma$ with $\dot{\gamma}$) is shown in Fig.~\ref{fig:6}(a), for a wide range of $a$ values, and the corresponding viscosity $\eta(\dot{\gamma})$ is shown in Fig.~\ref{fig:6}(b). 

For reference, we  show the flow curve for the non-breathing passive state, i.e. $a=0$. Since we are working at a packing fraction which is above $\phi_J$, the passive assembly behaves as a yield stress fluid, i.e. there exists a finite yield stress $\sigma_y$. The yield stress can be estimated via the Herschel-Bulkley fit $\sigma_{xy}=\sigma_y+A\dot{\gamma}^\beta$, where $A$ is a constant and $\beta$ is the Herschel-Bulkley exponent. 

Even when breathing is present, the shape of the curves as $\dot{\gamma} \rightarrow 0$ remains similar, i.e. the response is that of a yield stress fluid. Only when $a$ becomes larger than $a_c$, the curvature of the flow curve as $\dot{\gamma} \rightarrow 0$ changes and the viscosity
data shows a regime of constant $\eta$ emerging at small shear-rates, implying that for large breathing amplitudes, the unsheared state is a Newtonian fluid.
In the regime $a < a_c$, we again fit the flow curve data with the HB function. The variation of $\sigma_y$ with $a$ is shown in Fig.~\ref{fig:6}(c); we observe that the yield stress decreases with increasing breathing amplitude; i.e. the material softens as the oscillations increase in magnitude. Eventually, the rigidity is lost when $\sigma_y$ vanishes and we estimate the vanishing threshold by fitting the data with $(a^{\star}-a)^\zeta$ from 
where we obtain $a^{\star}=0.0506$ and $\zeta \approx 0.39$. We note that our estimate for the threshold amplitude for the loss of rigidity via the
probing of the steady state rheological response matches reasonably well, albeit slightly higher, with the threshold $a_c=0.0498$ estimated in Ref.\cite{tjhung2017discontinuous} from the dynamical data.  Further, in Fig.~\ref{fig:6}(d), we plot the shear-rate dependence of $\delta{\sigma_{xy}}=\sigma_{xy}-\sigma_c$ and we obtain a data collapse for all values of $a$ which are less than $a^{\star}$. This indicates that the $\delta{\sigma_{xy}}(\dot{\gamma})$ is the same for all $a<a^{\star}$, or in other words there is no change in the shear-thinning behaviour with increased oscillations, and also the data indicates that $\delta{\sigma_{xy}}$ vanishes as $\dot{\gamma}^{1/2}$. 

In addition to these shear-rate-controlled measurements, we further examine the response to constant applied stress in \ref{app:creep_response}. The observed creep behavior provides complementary evidence for activity-induced fluidization and supports the continuous softening of the material near the critical amplitude.

\begin{figure}
    \centering
    \includegraphics[width=0.85\linewidth]{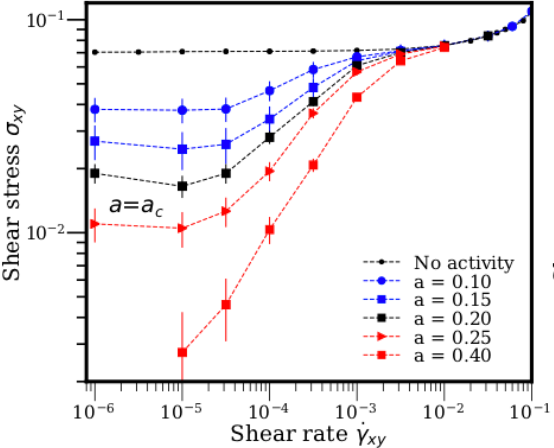}
    \caption{{\bf Mesoscale model.} Rheological flow curves for different breathing amplitudes.}
    \label{fig:mesorheo}
\end{figure}

%\begin{figure}[h]
%  \includegraphics[scale=0.4]{xfreq.pdf}
%  \caption{{\bf Microscopic model.} Frequency dependence of
%  the rheology. Data shown for $a=0.04$ (in green) and $0.07$ (magenta) for two different time periods, viz. 820 (in filled symbols) and 82 (in open symbols).}
%    \label{fig:7}
%\end{figure}

\subsubsection{Mesoscale model}

We show the mesoscale data for rheological response in Fig.~\ref{fig:mesorheo}. While the rheology remains largely unaffected by the activity at high shear rates, as evidenced by the collapse of the flow curves in this regime, we observe a pronounced decrease in shear stress with increasing activity amplitude at low to intermediate shear rates. This softening is indicative of a fluidization process driven purely by internal size fluctuations, closely mirroring the behavior seen in the microscopic simulations shown in Fig.~\ref{fig:6}.

Notably, the transition from a yield stress fluid to a more fluid-like response occurs without any external forcing. This highlights the ability of internal active deformations to reorganize stress fields throughout the system. The mesoscale model captures this behavior by coupling local active dilations or contractions to stress redistribution via an elastic propagator, which reduces the effective stress required to trigger plastic events.

Furthermore, the consistency between mesoscale and microscopic models supports the robustness of this internal activation mechanism. It suggests that the essential ingredients for activity-induced fluidization are the local injection of strain and the resulting long-range elastic coupling. This agreement validates the modeling approach and opens the door to future studies involving spatially heterogeneous or temporally modulated activity fields in soft glassy systems.

\section{Conclusions}
In this work, we have shown that active breathing, periodic size oscillations of soft particles, can drive an internally induced fluidization transition in dense amorphous assemblies. Using both microscopic simulations and a mesoscale tensorial elasto-plastic model, we demonstrate that this purely internal driving leads to a spontaneous unjamming transition beyond a critical amplitude. The transition is characterized by a sharp increase in plastic activity and a concurrent reduction in yield stress, ultimately resulting in Newtonian flow behavior at large activity.

These findings complement and extend earlier studies that identified size fluctuations as a key feature in confluent cellular monolayers \cite{zehnder2015cell} and demonstrated their potential to trigger unjamming transitions in simplified models \cite{tjhung2017discontinuous, li2025fluidization}. While previous work by Tjhung and Berthier \cite{tjhung2017discontinuous} established that breathing alone can fluidize a jammed configuration, our study goes further by systematically probing both the yielding transition and the full rheological response under external shear. We show that the reduction in yield stress with increasing activity amplitude follows a continuous trend and vanishes near a well-defined critical point, similar in spirit to observations of rigidity loss in sheared amorphous solids \cite{bonn2015yield, jaiswal2016mechanical}.

Furthermore, we demonstrate that the emergence of a constant-viscosity plateau at small shear rates marks a qualitative change in rheology from shear-thinning to Newtonian flow,  as the system transitions from a yield-stress material to a fluidized state. This behavior echoes the softening observed in driven particulate systems subjected to mechanical oscillations or external vibrations \cite{knowlton2014microscopic}, but here the drive originates internally and uniformly across the system. Our mesoscale model captures these features and offers insights into how local stress propagates from actively deforming sites, consistent with elasto-plastic descriptions of stress redistribution in amorphous solids \cite{review}. Our mesoscale model complements recent elastoplastic formulations developed for active glasses \cite{ghosh2025elastoplastic}, but differs by focusing on spatially distributed periodic activity and its coupling to macroscopic rheology.

A key distinction between our microscopic and mesoscale approaches lies in the nature of the imposed activity: the mesoscale model uses idealized sinusoidal forcing, while the microscopic model results in periodic but not strictly sinusoidal deformations. Despite this, both models yield consistent qualitative trends, reinforcing the notion that periodic internal driving rather than the precise waveform is sufficient to trigger the observed behavior.

In addition, we observe that at low activity amplitudes, plastic events remain synchronized with the breathing cycle and are spatially localized. At higher amplitudes, rearrangements span multiple cycles and become more sensitive to local structural heterogeneities. This mirrors the behavior of cyclically sheared amorphous materials, where transitions from reversible to irreversible dynamics and the emergence of multi-periodic or chaotic responses are well documented \cite{fiocco2013oscillatory, regev2013onset}.

Altogether, our results position active size fluctuations as a robust internal mechanism for mechanical regulation in dense disordered materials. They not only reinforce recent observations in biological systems where size fluctuations and internal contractility drive tissue remodeling but also suggest a general principle by which active materials can modulate their rigidity without external forcing. Future work should explore how these principles extend to more complex or heterogeneous environments, and how tuning the spectrum or spatial distribution of activity might be used to engineer programmable mechanical behavior in synthetic active matter systems.

%[\textcolor{red}{Work to be done.}] Regarding the similarity with cyclically sheared systems, we note that, like cyclically sheared systems, we see that once “trained” at a certain breathing amplitude, the system retains the memory of said amplitude (Fig. \ref{fig:5}).

%\newpage

\section{Acknowledgements}

The authors acknowledge funding for this project via the CEFIPRA Project 5604-1.

\appendix

\section{\label{app:creep_response} Creep response under constant stress}

\begin{figure}[h]
% \centering
 %   \includegraphics[scale=0.8]{gamma_t_s-0.0010_a.png}
%    \includegraphics[scale=0.35]{12.png}\\
        \includegraphics[scale=0.35]{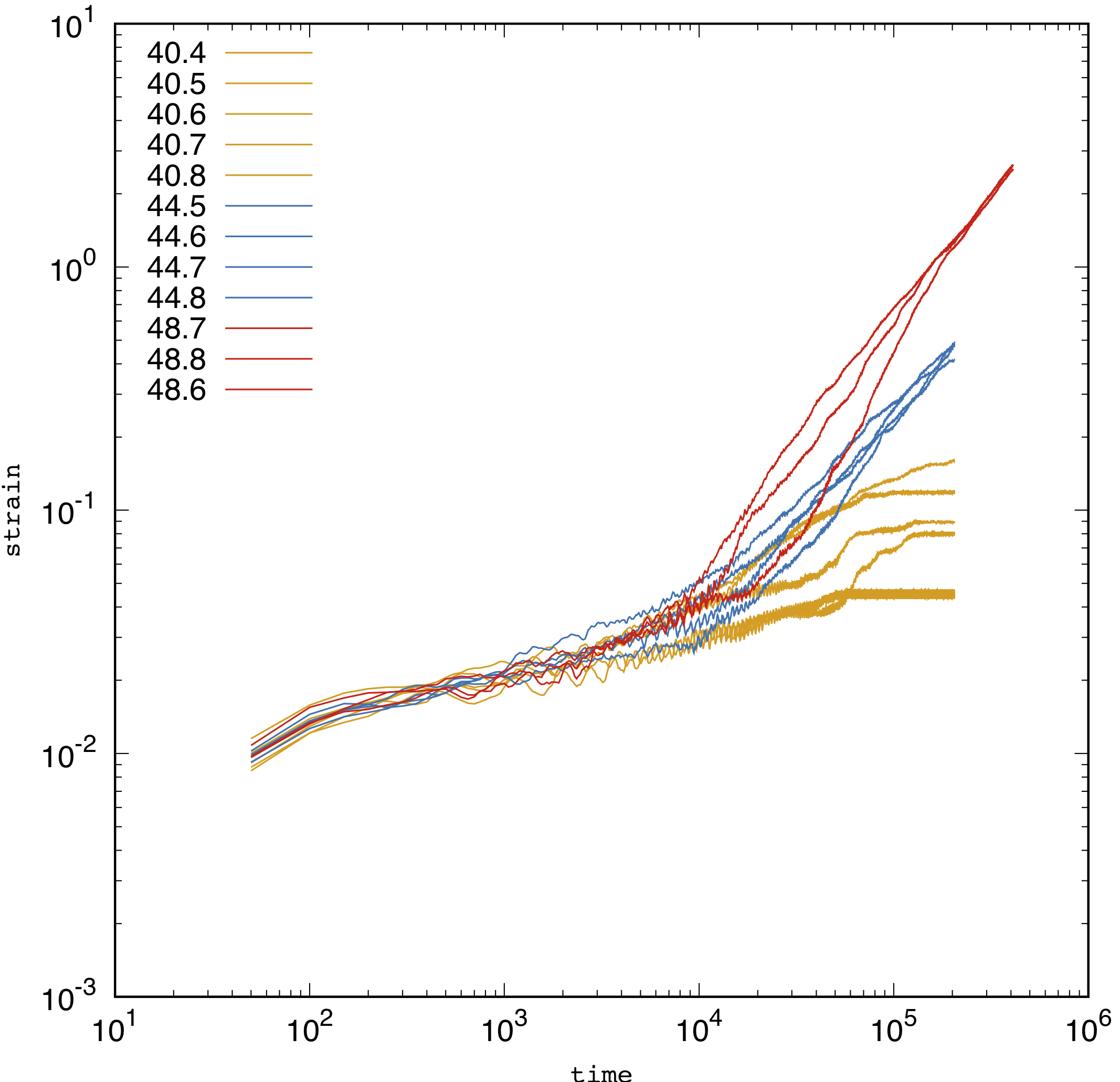}
    \caption{{\bf Microscopic model.} Creep curves for imposed stresses of 0.0015 for breathing amplitudes of $0.040, 0.044, 0.048$.}
    \label{fig:8}
\end{figure}

To complement the rheological characterization under imposed shear rates, we also examine the system's response to a constant applied shear stress~\cite{cabriolu2019precursors}. In Fig.~\ref{fig:8}, we show creep curves for different values of the breathing amplitude $a$, all subjected to the same external stress $\sigma_0 = 0.0015$. At low activity ($a = 0.040$), the strain evolves slowly and saturates, indicating a jammed response. As the activity increases, the system begins to flow more readily, with the strain growing steadily in time, consistent with fluidization. These results reinforce the picture that internal active deformations reduce the mechanical resistance of the system and lower the threshold for flow onset, even in the absence of thermal fluctuations.

\section{\label{app:active_response} Calculation of the linear elastic response to dilating particles}

\begin{figure}
    \centering
    \includegraphics[width = 0.3\columnwidth, clip]{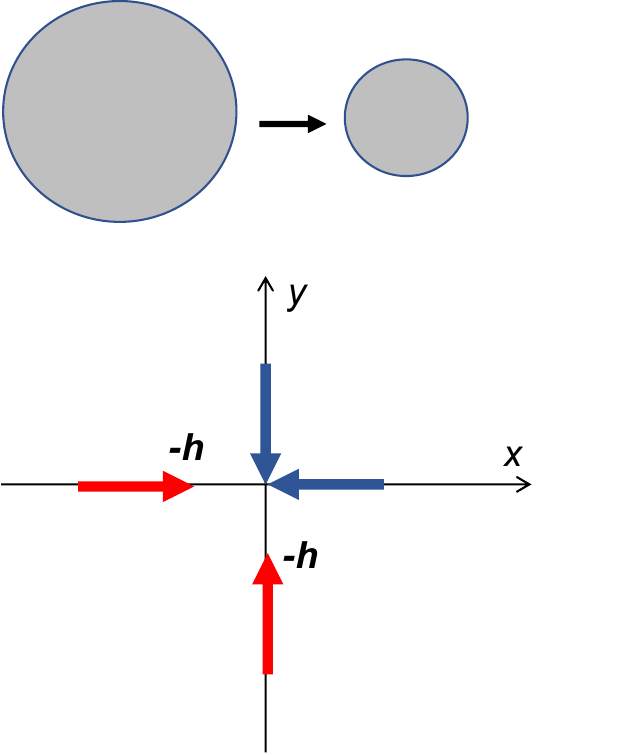}
    \caption[Force dipoles to model a dilating or contracting inclusion]{\textbf{Schematic of two force dipoles to model a dilating or contracting inclusion}}. 
    \label{fig:dipole}
\end{figure}

The dilation (or contraction) of the active particle is modeled by two orthogonal pairs of forces. Each pair consists of two forces of equal magnitude $f_0$ but opposite directions acting at two points separated by a distance $h$, as illustrated in figure \ref{fig:dipole}.
Such a pair of force exerts no net force on the material.
We assume that one force pair acts in the $x$-direction and the other one in the $y$-direction and that the defect is centered at the origin. The total force density is given by:
\begin{eqnarray}
\vec{f}(\vec{r}) &=& -f_0 \delta(\vec{r})\vec{e_x} + f_0 \delta(\vec{r}-h\vec{e_x}) \vec{e_x} \nonumber\\ &&- f_0 \delta(\vec{r})\vec{e_y} + f_0 \delta(\vec{r}-h\vec{e_y}) \vec{e_y} 
\end{eqnarray}
where $\vec{e_x}$ and  $\vec{e_x}$ are unit vectors in the $x$- and $y$-direction and $\delta(\vec{r})$ is the delta Dirac function in two dimensions.
%%%%%%%%%%%%%%%%%%%%%%%%%%%%%%%%%%%%%%%%%%%%%%
%%%%%%%%%%%%%%%%%%%%%%%%%%%%%%%%%%%%%%%%%%%%%%%
By taking the limit $h \rightarrow 0$ (ensuring that $f_0 \rightarrow \infty$ so $f_0 h$ remains finite), we get:
\begin{equation}
\vec{f}(\vec{r}) = -f_0 h \vec{\nabla} \delta(\vec{r})
\end{equation}
%We rewrite the force using indicial notations:
%\begin{equation}
%f_\alpha(x_\beta) = -f_0 h \frac{\partial \delta(x_\beta) }{\partial x_\alpha} 
%\end{equation}
%
The Navier equation describing the static state of a material reads
\begin{equation}
(\lambda + \mu) \vec{\nabla} (\vec{\nabla} \cdot \vec{u} ) + \mu \Delta \vec{u} = - \vec{f}
\end{equation}
Taking the divergence on both sides of the above equation and using the expression for $\vec{f}$ yields 
\begin{equation}
\Delta (\lambda + 2\mu) (\vec{\nabla} \cdot \vec{u} ) = Fh \Delta \delta(\vec{r})
\end{equation}
To find the displacement field obeying the above equation, one can then write the displacement field as deriving from a scalar potential $\phi$: $\vec{u} = \vec{\nabla} \phi$ (see e.g., \cite{lechner2009point} for a detailed derivation). 
This is analogous to the Poisson equation for a punctual charge in electrostatics. A possible solution of the equation $\Delta K = \delta(r)$ is $K(r) = \log(r)/2\pi$ thus leading to :
$\phi (\vec{r}) = \alpha \log(\vec{r})$ with $\alpha =  \frac{Fh}{(\lambda + 2\mu) }$ (for a vanishing displacement field at infinity).  
We finally get for an isotropic response for the displacement field:
\begin{equation}
\vec{u}(\vec{r}) = \alpha \frac{\vec{r}}{r^2}
\end{equation}
We re-write the two components of the displacement field in polar coordinates ($r,\theta$):

\begin{equation}
u_x(r,\theta) = \alpha \frac{\cos{\theta}}{r} \quad \mathrm{,} \quad u_y(r,\theta) = \alpha  \frac{\sin{\theta}}{r}
\end{equation}
The stress field in response to a dilating inclusion then reads:
\begin{equation}
\sigma_{ij} = 2 \mu \varepsilon_{ij} = \mu \left( \frac{\partial u_i}{\partial x_j} + \frac{\partial u_j}{\partial x_i} \right)
\end{equation}
And we get for the two shear components: 
\begin{eqnarray}
\sigma_{xy}(r,\theta) &=& - \alpha \mu \frac{\sin{2\theta}}{r^2} \quad \mathrm{,} \\  \frac{\sigma_{xx}-\sigma_{yy}}{2}(r,\theta) &=&  - \alpha \mu \frac{\cos{2\theta}}{r^2} 
%\sigma_{xy} = 4 \alpha \mu \frac{xy}{(x^2+y^2)^2} 
\end{eqnarray}
while the pressure response is null: $- \frac{\sigma_{xx}+\sigma_{yy}}{2}(r,\theta) = 0$.
%\begin{equation}
%\sigma_{xx} = \frac{\sigma_{xx}-\sigma_{yy}}{2} =  2  \alpha \mu %\frac{y^2-x^2}{(x^2+y^2)^2} 
%\end{equation}

%\bibliography{references}

\end{document}